\documentclass[pra,twocolumn,showpacs,superscriptaddress,preprintnumbers,amsmath,amssymb,aps]{revtex4-1}
\usepackage{mathrsfs}
\usepackage{amsmath}
\usepackage{amssymb}
\usepackage{graphicx}
\usepackage{dcolumn}
\usepackage{bm}
\usepackage{subfigure}
\usepackage{color}
\usepackage{ifpdf}
\usepackage{verbatim}
\usepackage{bbm}

\usepackage[
pdfstartview=FitH,%
bookmarks=true,%
bookmarksopen=true,%
breaklinks=true,%
colorlinks=true,%
linkcolor=blue,anchorcolor=blue,%
citecolor=blue,filecolor=blue,%
menucolor=blue,%
urlcolor=blue]{hyperref}
\usepackage{float}
\usepackage[toc,page,title,titletoc,header]{appendix}

\def\be{\begin{equation}}
\def\ee{\end{equation}}
\def\bea{\begin{eqnarray}}
\def\eea{\end{eqnarray}}
\def\nn{\nonumber}

\begin{document}

\title{Simulation of nodal-line semimetal in amplitude-shaken optical lattices}
\author{Tanji Zhou}
\affiliation{State Key Laboratory of Advanced Optical Communication Systems and Networks, Department of Electronics, Peking University, Beijing 100871, China}
\affiliation{School of Physics, Peking University, Beijing 100871, China}
\author{Zhongcheng Yu}
\affiliation{State Key Laboratory of Advanced Optical Communication Systems and Networks, Department of Electronics, Peking University, Beijing 100871, China}
\affiliation{School of Physics, Peking University, Beijing 100871, China}
\author{Zhihan Li}
\affiliation{State Key Laboratory of Advanced Optical Communication Systems and Networks, Department of Electronics, Peking University, Beijing 100871, China}
\author{Xuzong Chen}
\affiliation{State Key Laboratory of Advanced Optical Communication Systems and Networks, Department of Electronics, Peking University, Beijing 100871, China}
\author{Xiaoji Zhou}\email{xjzhou@pku.edu.cn}
\affiliation{State Key Laboratory of Advanced Optical Communication Systems and Networks, Department of Electronics, Peking University, Beijing 100871, China}
\affiliation{Collaborative Innovation Center of Extreme Optics, Shanxi University, Taiyuan, Shanxi 030006, China}
\date{\today}

\begin{abstract}
As the research of topological semimetals develops, semimetals with nodal-line ring come into people's vision as a new platform for studying electronic topology.
We propose a method using ultracold atoms in a two-dimensional amplitude-shaken bipartite hexagonal optical lattice to simulate the nodal-line semimetal, which can be achieved in the experiment by attaching one triangular optical lattice to a hexagonal optical lattice and periodically modulating the intensity and position of the triangular lattice. 
By adjusting the shaking frequency and well depth difference of the hexagonal optical lattice, a transformation from Dirac semimetal to the nodal-line semimetal is observed in our system. This transformation can be demonstrated by the Berry phase and Berry curvature, which guides the measurement.
Furthermore, the amplitude shaking and $C_3$ symmetry generate a time-reversal-symmetry-unstable mode, and the proportion
of the mode and the trivial mode of the hexagonal lattice controls the transformation. 
This proposal paves a way to study nodal-line semimetals in two-dimensional systems, and it contribute to the application of nodal-line semimetals. 
\end{abstract}


\maketitle

\section{Introduction}
Since Hermann Weyl found a massless solution to the Dirac equation \cite{Weyl1929} in 1929, the research of Weyl fermion has attracted intensive attention. Last century, the neutrino was considered as a strong candidate for Weyl fermion, until the observation of neutrino oscillation shows that neutrino is not massless \cite{PhysRevLett.87.071301}. Recently a breakthrough about Weyl fermion has been achieved -- a gapless semimetal, named Weyl semimetal, was realized in photonics, condensed matter, and cold atoms \cite{Lu2013,PhysRevLett.111.027201,Huang2015,HE2018440,Noh2017}. In Weyl semimetal, low-energy excited electrons, at Weyl points, behave as Weyl fermion \cite{Lu622,Xu613,doi:10.1146/annurev-conmatphys-031016-025458}. Two Weyl points with opposite chirality can be considered as splitting from a Dirac point by breaking time-reversal symmetry or inversion symmetry \cite{Chen2015,doi:10.1146/annurev-conmatphys-031016-025458}.
During the research, a new topological semimetal was discovered, which is called nodal-line semimetal \cite{PhysRevB.84.235126}.

As one kind of nontrivial topological semimetals, the nodal-line semimetal is different from Weyl semimetal and Dirac semimetal \cite{,PhysRevLett.115.036807,Hirayama2017,PhysRevB.93.205132,PhysRevLett.121.036401}.
In Weyl semimetal and Dirac semimetal, the bands cross at points. However, the valence and conduction bands cross and form a ring-shaped nodal line in nodal-line semimetal \cite{PhysRevLett.115.026403,PhysRevLett.115.036806,PhysRevB.95.045136}.In bulk states, the unusual electromagnetic and transport response of nodal-line semimetals have attracted intense attention \cite{PhysRevB.81.195431,PhysRevB.93.045201,PhysRevB.94.195123,PhysRevB.96.245101}. In terms of surface states, Weyl semimetal has the Fermi arc, which is the intersection of the 2D surface of the Brillouin zone and stretches between two Weyl points. While in nodal-line semimetal, the so-called “drumhead”-like states are the flat bands nestled inside of the circle on the surface projected from bulk nodal-line\cite{PhysRevB.92.045108,PhysRevB.93.121113}, which is different from Fermi arc and may demonstrate the presence of interactions\cite{PhysRevB.96.245101}. 
Further, nodal-line semimetal is expected to have high fermion density at nodal-line ring due to the crossing of two bands, which may contribute to filling the gap between fundamental physics of topological materials and practical applications in quantum devices\cite{PhysRevLett.117.016602}.

\begin{figure*}[htp]
	\includegraphics[width=0.95\textwidth]{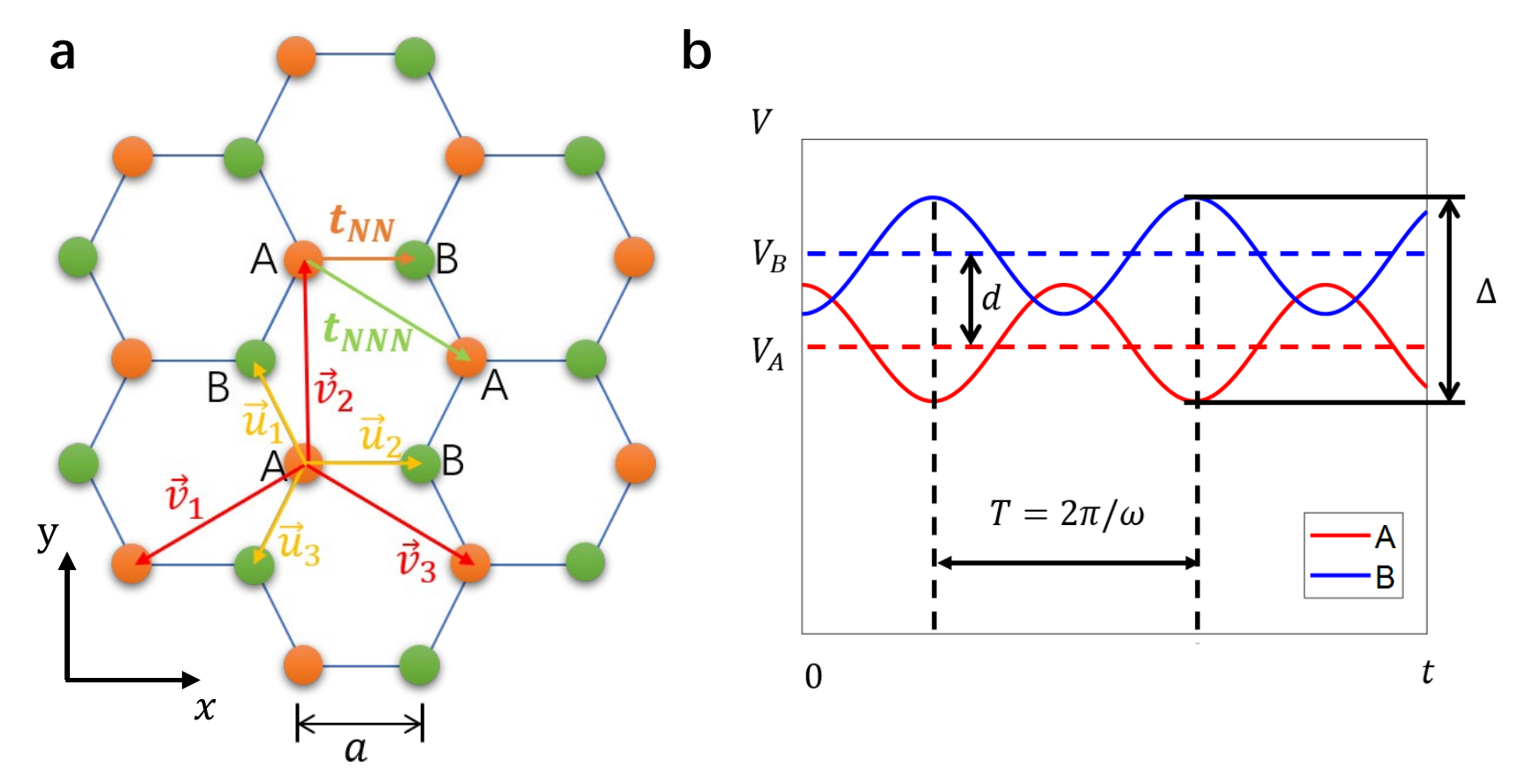}
	\caption{\textbf{Schematic diagram of 2D bipartite hexagonal lattice.} (a) The lattice depth at point A and B of the bipartite hexagonal lattice are different and can be modulated periodically. $\vec{v_j}$ are lattice vectors, $\vec{u_j}$ are vectors connecting nearest-neighbour points ($j=$1,2,3). $a$ is the distance between nearest neighbor points. (b) We change the amplitude shaking of the lattice depth at A and B as the form of cosine function with a phase difference $\pi$, where the orange solid line and green dotted line denote points A and B, respectively. $d$ is the difference of the average lattice depth between points A and B. $\omega$ and $T$ denote the frequency and period of shaking, respectively. $V_A$ and $V_B$ are the lattice depth of trivial hexagonal optical lattice, and $\delta=V_B-V_A$ is the well difference between points A and B. $\Delta$ is the shaken amplitude of well depth.
	}\label{fig:fig1}
\end{figure*}

On the other hand, the development of the artificial gauge field has paved the way for simulating topological materials by using ultracold atoms \cite{RevModPhys.83.1523,PhysRevLett.109.145301,PhysRevLett.111.120402,PhysRevLett.112.086401,PhysRevLett.113.045303,Jotzu2014,doi:10.1063/1.5040669,Songeaao4748,Guo19}.
Furthermore, due to their high controllability, ultracold atoms in optical lattices are widely used to simulate unique topological phenomena, including the measurement of the second Chern number \cite{Lohse2018}, the observation of topologically protected edge states \cite{Stuhl1514,Leder2016}, and the chiral interaction of Weyl semimetals \cite{PhysRevLett.114.225301,Zhang_2016,PhysRevA.95.033629}. With the development of corresponding technologies, recently, 3D spin-orbit coupled ultracold atoms in optical lattice have been used in the simulation of nodal-line semimetal \cite{Song2019}. In condensed matter systems, due to its complexity, detection of the nodal-line ring is always influenced by other bulk bands \cite{PhysRevB.96.245101,PhysRevB.92.045108,PhysRevB.93.121113}. In comparison, optical lattices are more controllable and can greatly remove influence from other bulk bands to directly observe nodal-line ring.

In this paper, we propose a method to simulate nodal-line semimetal based on ultracold atoms in an amplitude-shaken bipartite hexagonal optical lattice. By periodically changing the lattice depth of the bipartite hexagonal optical lattice, a time-reversal-symmetry-unstable mode is imported into the system, which corresponds to nodal-line phase. As the frequency and shaking amplitude change, the energy spectrum of the optical lattice transforms from Dirac semimetal to nodal-line semimetal.
Then we discuss symmetry of the system which causes the transformation and analyze the topological characteristics of the shaking optical lattice. The Berry curvature and Berry phase are demonstrated in the transformation process, which provides a method to detect the transformation in further experiments. 
Finally, the influence of the well depth difference of the hexagonal optical lattice on the nodal-line phase is specifically discussed, which indicates the effect of crystalline symmetries on the nodal-line phase.
Combining with the advantages of ultracold atoms, it is likely to pioneer a new approach to nodal-line semimetal in two-dimensional systems. Moreover, it may serve as an important basis for future studies of the symmetry of nodal-line semimetals. 
Different from existing works \cite{Song2019}, our system has three distinguishing features: (1) We used amplitude-shaken optical lattice to simulate nodal-line semimetal, which is effective and easy to be implemented, while a great majority of previous methods use spin-orbit coupling(SOC) in an optical lattice. (2) Our method can simulate a two-dimensional nodal-line semimetal, while the existing works mainly focus on three-dimensional systems.   (3) Band structure in our system takes the form of semiconductor, where the conduction and valence bands touch at nodal-line ring without crossing each other, while existing works about nodal-line semimetal are mainly semimetal form, which means the existence of an overlap between the bottom of the conduction band and the top of the valence band. This difference directly influences the distribution of the density of states, which may cause unusual electromagnetic and transport response. Here we use the common name ``nodal-line semimetal'' to describe our system, while the name ``nodal-line semiconductor'' may be more suitable in practical terms.

The remainder of this paper is organized as follows. In Sec.II, we introduce a particular model which is called the amplitude-shaken bipartite hexagonal optical lattice and propose its feasible experimental scheme. In Sec.III, we describe the calculation to get effective Hamiltonian and derive the energy dispersion of our system. The band structure during the transformation process from Dirac semimetal to nodal-line semimetal is shown and explained by symmetry in
Sec.IV. The discussion about topological properties by calculating Berry curvature and Berry phase in a certain area and the study of nodal-line phase with well depth difference are shown in Sec.V. Then we give a conclusion in Sec.VI.

\section{Model description and feasible experimental scheme}
\subsection{Model description}

In this model, the amplitude shaking is applied to a trivial honeycomb optical lattice, and the relationship between well depth $V_i$ and time $t$ is shown in Fig.\ref{fig:fig1}, where $i=A,B$. In these two figures, A and B are different points which are inequivalent in honeycomb lattice structure with a difference of well depth $d$. $t_{NN}$ and $t_{NNN}$ denote the nearest-neighbor hopping coefficient and next-nearest-neighbor hopping coefficient, respectively. And $\vec{v_j}$ are lattice vectors, where $\vec{v_1}=(-3a/2,-\sqrt{3}a/2),\vec{v_2}=(0,\sqrt{3}a)$ and $\vec{v_3}=(3a/2,-\sqrt{3}a/2)$. $\vec{u_j}$ are vectors connecting nearest-neighbour points, where $\vec{u_1}=(-\sqrt{3}a/2,a/2),\vec{u_2}=(a,0)$ and $\vec{u_3}=(-\sqrt{3}a/2,-a/2)$. $a=2\sqrt{3}\lambda/9$ is the distance between nearest neighbor points, which is the side length of smallest hexagon, where $\lambda$ is the wavelength of the lattice laser beam.

The amplitude-shaken model is shown in Fig.\ref{fig:fig1}(b) .The lattice depths of points A and B shake periodically in the form of the cosine function at a fixed phase difference $\pi$. In the physical image, the hexagonal lattice can be divided into two sets of amplitude shaking triangular lattices, the amplitude of each shakes with a phase difference $\pi$ between them. In the figure, $\omega$ is the angular frequency of periodically shaking, and $\Delta$ is the shaking amplitude, which is small enough to be considered as perturbation comparing with $V_{A,B}$, where $V_{A,B}$ is the lattice depth of point A and B.

Considering ultracold atoms, we neglect the weak atomic interactions as demonstrated in usual experiments  \cite{ARIMONDO2012515}, and start with a single atom model in the honeycomb lattice. The Hamiltonian $\hat{H}$ of the system can be written as the zeroth-order Hamiltonian of the honeycomb lattice $\hat{H_0}$ adding a first-order perturbation $\hat{H_1}$ caused by amplitude shaking
\begin{eqnarray}
\label{H}
\hat{H}=\hat{H_0}+\hat{H_1} 
\end{eqnarray}

According to the single-particle two-band tight-binding model, $\hat{H_0}$ can be written as:
\begin{eqnarray}
\label{H 0}
\hat{H_0}=\sum_{i}V_{i0} c_i^\dag c_i +\sum_{i \not = j }t'_{ij} c_i^\dag c_j
\end{eqnarray}
where $i$ represents each node of honeycomb lattice,  $V_{i0}$ equals to $V_{A0}$ and $V_{B0}$ for A and B site respectively, with a difference $V_{B0}-V_{A0}=d$. The operators $c_i^\dagger$ and $c_i$ denote the creation and annihilation operator at node $i$. $t'_{ij}$ denotes the hopping coefficient between point $i$ and $j$. And $\sum_{i \not = j }$ denotes summation over all pairs of different points.

Next, we give the expression $\hat{H_1}$ when the shaking of the lattice depth
takes the following form:
\begin{eqnarray}
\label{depth of A and B}
V_i(t)=V_{i0}+\chi(i)\frac{\Delta}{2} \cos(\omega t)
\end{eqnarray}
where $\omega$ denotes the frequency of shaking. $\chi(i)$ at point A equals to $1$, while at point B equals to $-1$. So $\hat{H_1}$ reads
\begin{eqnarray}
\label{H1}
\hat{H_1}=\sum_{i}\chi(i)\frac{\Delta}{2} \cos(\omega t)c_i^\dag c_i
+\sum_{i \not = j } \delta t_{ij}(t) c_i^\dag c_j
\end{eqnarray}
where $\delta t_{ij}(t)$ is the change of hopping coefficient due to amplitude shaking.

\begin{figure}
	\includegraphics[width=0.5\textwidth]{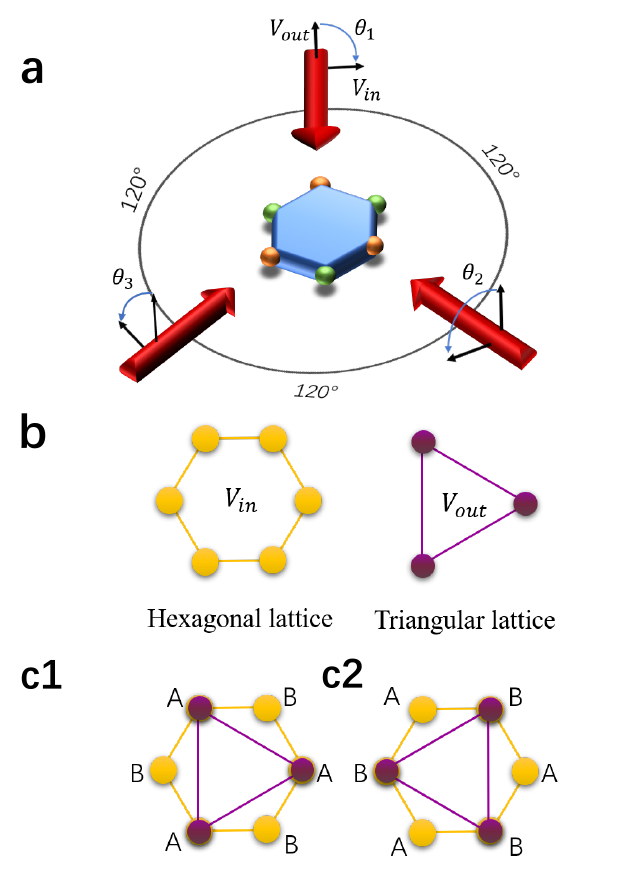}
	\caption{\textbf{The proposed experimental scheme diagram.} (a) Using 3 intersecting elliptical polarized laser beams with the enclosing angle of 120 degrees, the bipartite hexagonal is formed. Each laser beam is formed by combining two linearly polarized laser beams with polarization directions in the lattice plane ($V_{in}$) and perpendicular to the lattice plane ($V_{out}$). (b) Hexagonal lattice corresponds to $V_{out}$ and triangular lattice corresponds to $V_{in}$. (c) Through changing the relative phases of three lasers, two different methods of overlying hexagonal lattice and triangular lattice are formed. In situation (c1), the lattice depth of point B is higher than A. When the intensity of the triangular reduce to 0, we move the triangular lattice instantaneously in order to let the lattice depth of point A become higher than B, as is the situation in (c2).
	}\label{fig:fig1_2}
\end{figure}

Therefore, the Hamiltonian $\hat{H}$ of the system can be written as:
\begin{eqnarray}
\label{H111}
\hat{H}=\sum_{i}V_i(t) c_i^\dag c_i +\sum_{i \not = j }t_{ij} c_i^\dag c_j
\end{eqnarray}
where $t_{ij}(t)=t'_{ij}+ \delta t_{ij}(t)$ is the hopping coefficient between point $i$ and $j$ as a function of time. In subsequent calculations, we only keep the nearest and the next-nearest hopping coefficient. 

During the shaking process, since each pair of next-nearest sites belong to the same category of points, which follow the identical rule of changing in the lattice depth, the difference in well depth between them is always zero. Thus we can consider the next-nearest hopping coefficient $t_{NNN}$ as a constant. And the nearest hopping coefficient $t_{NN}$ equals to one constant $t_0$ adding one term which is proportional to the lattice depth difference between point A and B (See Appendix A), so we define $t_{NN}$ as
\begin{eqnarray}
\label{tnn}
t_{NN}=t_0+t_1 \cos\omega t
\end{eqnarray}
where $t_1$ is a constant which has the unit of energy.

\subsection{Feasible experimental scheme}
The above 2D amplitude shaking model can be constructed with the following experimental scheme. This scheme is easy to be expanded to the 3D system by adding a laser beam perpendicular to the lattice plane \cite{Jin_2019}. As shown in Fig.\ref{fig:fig1_2}(a), we use three elliptical polarized laser beams with an enclosing angle of 120$^\circ$ to each other to form a bipartite optical hexagonal lattice, which has been demonstrated in recent works \cite{Flschner1091,jin2019dynamical}. The total potential of optical lattice is written as
\begin{eqnarray}
\label{V_lattice}
V(\vec{r})=-V_{\rm out}\sum_{i',j'}{\rm cos}\left[(\vec{k_{i'}}-\vec{k_{j'}})\cdot \vec{r}-(\theta_{i'}-\theta_{j'})\right] \\
+\frac{1}{2}V_{\rm in}\sum_{i',j'}{\rm cos}\left[(\vec{k_{i'}}-\vec{k_{j'}})\cdot \vec{r}\right]\nn
\end{eqnarray}
where $i',j' =1,2,3$ represent three directions of wave vectors of laser, and $\vec{k_1}= (\sqrt{3}\pi,-\pi)/\lambda$, $\vec{k_2}= (-\sqrt{3}\pi,-\pi)/\lambda$, $\vec{k_3}= (0,2\pi)/\lambda$ are three wavevectors. $\lambda$ is the wavelength of the laser beam, $V_{\rm out}$ and $V_{\rm in}$ denote the components perpendicular and parallel to the lattice plane, which can be controlled respectively. And the three angles $\theta_1,\theta_2,\theta_3$ represent the relative phases of the elliptical polarization of the laser beams.

It can be considered as a triangular optical lattice adding to a hexagonal one, where $V_{\rm out}$ corresponds to the triangular optical lattice and $V_{\rm in}$ corresponds to the hexagonal optical lattice, as shown in Fig.\ref{fig:fig1_2}(b). Through adjusting the phase $\theta_i$, the position of the triangular optical lattice can be modulated. At first, the position of two sets of lattices is as shown in Fig.\ref{fig:fig1_2}(c1), where $\theta_{1,2,3}=(2\pi/3,4\pi/3,0)$.
So, the lattice depth at point A is the superposition of potential wells of two sets of optical lattices, while at point B the lattice depth is the superposition of the potential well of hexagonal optical lattice and the potential barrier of triangular one.
Then through modulating $V_{\rm out}$ in the form of the cosine function, the lattice depth at points A and B can change in the form of function $|cos|$. But in this way, $V_B$ is always higher than $V_A$. So, in order to get the shaking curve as shown in Fig.\ref{fig:fig1}(b), we need to adjust the position of triangular optical lattice to make the lattice depth at point A and B reverse, by modulating $\theta_{1,2,3}$ to $(2\pi/3,4\pi/3,0)$ when the intensity of triangular lattice reduces to 0. Fig.\ref{fig:fig1_2}(c2) shows the position of triangular optical lattice at this situation. The time of reversion, in theory, can be reached within several microseconds, and the shaking period in our protocol is around $500~\mu s$. As long as the time of reversion is short enough comparing with the oscillating period, the continuity of this process remains intact.

\section{Effective Hamiltonian and energy dispersion relation of the Floquet system}
In order to study the nodal-line semimetal, we first calculate the band structure of the system.
In the above section, we derive the Hamiltonian of the system as a function of time in Equation (\ref{H111}). In this section, we introduce the method to drive the effective Hamiltonian and get the band structure of the system, which can be considered as averaging the $\hat{H}$ over time.

To start with, we act an unitary transformation on $\hat{H}$. Define unitary operator $\hat{U}$:
\begin{eqnarray}
\label{U}
\hat{U}=\exp\left(\dfrac{\mathbbm{i}}{\hbar} \sum_{i}\int_{0}^{t}d\tau \cdot V_i(\tau)c_i^\dag c_i\right)
\end{eqnarray}
Through this unitary transform, we get $e$-index form $\hat{H'}$ (See Appendix B):
\begin{eqnarray}
\label{H tr}
\hat{H'}&&=\hat{U}\left(\hat{H}(t)-\mathbbm{i} \hbar\dfrac{\partial}{\partial t}\right)\hat{U}^\dag-\left(-\mathbbm{i} \hbar\dfrac{\partial}{\partial t}\right) \\
&&=-\dfrac{d}{2}\sum_i (a_i^\dag a_i-b_i^\dag b_i) -\sum_{i \not = j }t_{ij} e^{\mathbbm{i}z_{ij}\sin\omega t} c_i^\dag c_j \nn 
\end{eqnarray}
where $z_{ij}=\frac{\Delta}{2 \omega \hbar}[\chi(i)-\chi(j)]$ characterizes the responses of amplitude shaking at different sites. 

Next, the effective Hamiltonian can be obtained by using high-frequency expansion. We only keep up to first-order terms, and the effective Hamiltonian takes the form (More details see Appendix C):
\begin{eqnarray}
\label{H eff}
H_{eff}=H_{eff}^{(0)}+\dfrac{1}{\hbar\omega}H_{eff}^{(1)}=H_{f0}+\sum_{n=1}^{\infty} \frac{[H_{n},H_{-n}]}{n \hbar \omega}
\end{eqnarray}
where $H_{f0}$ is the zeroth-order term, and $H_{n}$ means the $e^{in\omega t}$ term Fourier expansion coefficients of Hamiltonian in Eq.(\ref{H tr}).
Then we get the kernel of effective Hamiltonian:
\begin{eqnarray}
\label{Heff}
&&\mathcal{H}_{eff}=\mathcal{H}_{eff,0}\hat{I}+\mathcal{H}_{eff,x}\hat{\sigma _x}+\mathcal{H}_{eff,y}\hat{\sigma _y}+\mathcal{H}_{eff,z}\hat{\sigma _z} \nn \\
&&
\end{eqnarray}
where
$$\mathcal{H}_{eff,0}=-2t_{NNN}\sum\limits_{j=1}^{3}\cos(\vec{k} \cdot \vec{v_j})$$
$$\mathcal{H}_{eff,x}=-t_0\mathcal{J}_0(\beta) \sum\limits_{j=1}^{3}\cos(\vec{k} \cdot \vec{u_j})$$
$$\mathcal{H}_{eff,y}=-t_0\mathcal{J}_0(\beta) \sum\limits_{j=1}^{3}\sin(\vec{k} \cdot \vec{u_j})$$
$$\mathcal{H}_{eff,z}=\dfrac{8t_0t_1}{\Delta} \mathcal{J}^2_1(\beta) \left(\sum\limits_{j=1}^{3}\cos(\vec{k} \cdot \vec{v_j})+\dfrac{3}{2}-\delta\right)
$$
 $\hat{\sigma _x},\hat{\sigma _y},\hat{\sigma _z}$ are Pauli matrices, and $\vec{k}$ is the wave vector of atomic state function. $\mathcal{J}_n$ means the $n{\rm th}$ order Bessel function of the first kind.
Here we define shaking factor $\beta$ and $\delta$ to describe the shaking:
\begin{eqnarray}
\label{beta}
\beta \equiv \frac{\Delta}{\hbar \omega}
\end{eqnarray}

\begin{eqnarray}
\label{delta}
\delta \equiv \dfrac{d/2}{8t_0t_1\mathcal{J}^2_1(\beta)/\Delta}
\end{eqnarray}

$\beta$ and $\delta $ are the main parameters affecting the effective Hamiltonian since they include shaking frequency $\omega$, shaking amplitude $\Delta$, and the difference of well depth $d$.
\begin{figure}
	\includegraphics[width=0.5\textwidth]{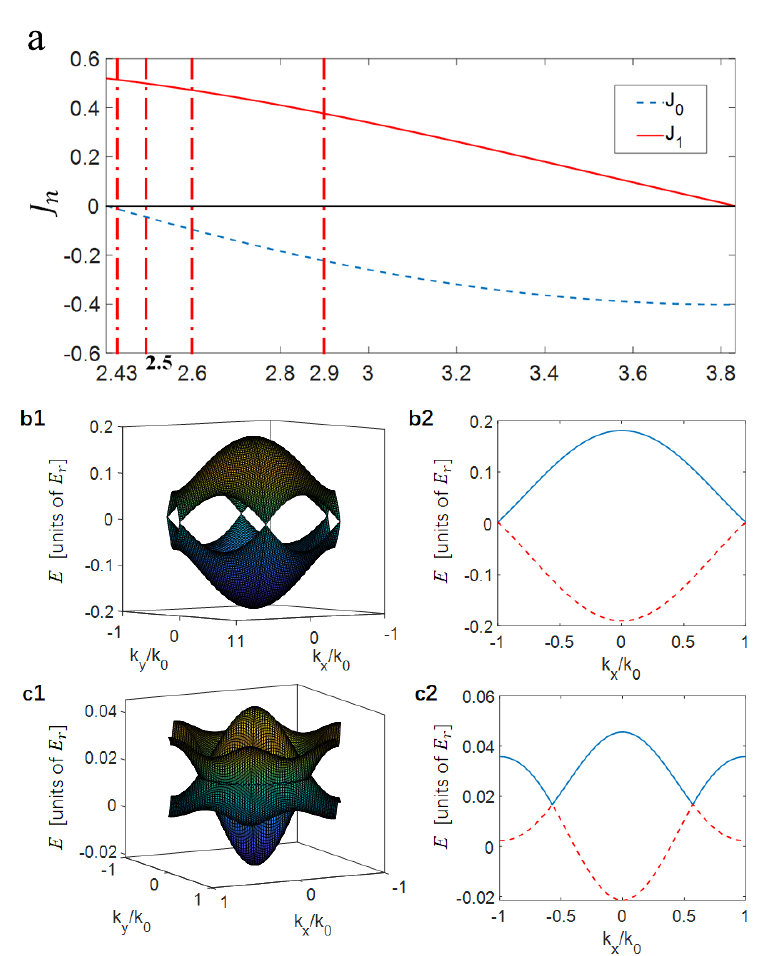}
\caption{\textbf{Band structure of the system.} (a) shows the variation curve of $\mathcal{J}_0(\beta)$ and $\mathcal{J}_1(\beta)$, when $\beta$ changes from one of zeros of zeroth-order Bessel  function $\mathcal{J}_0$ ($\beta=2.4048$) to one of zeros of first-order Bessel  function $\mathcal{J}_1$ ($\beta=3.8317$). And the four dot dashed red lines mark four characteristic points during the transformation, at $\beta$ = 2.43, 2.50, 2.60, 2.90, which is to show the main character of the transformation process later in Fig.\ref{fig:fig4}. (b) The band structure of Dirac semimetal. (b1) When $\mathcal{J}_1(\beta)=0$, the optical lattice performs as Dirac semimetal. (b2) is the sectional view of over-momentum zero corresponding to (b1), where $k_y$ has been set to zero and we focus on the change of energy with $k_x$. (c) The band structure of nodal-line semimetal. (c1) When $\mathcal{J}_0(\beta)=0$, the optical lattice performs as nodal-line semimetal. (c2) is the sectional view of over-momentum zero corresponding to (c1). In this situation, conduction band and valence band touch at ring $k_x^2+k_y^2=(\frac{4 \pi}{9})^2/a^2$. In the figure, $k_0=\frac{4\pi}{3\sqrt{3}a}$ is the distance from center of first Brillouin zone to the Dirac point.
	}\label{fig:fig2}
\end{figure}
Through solving eigenvalues of the kernel of effective Hamiltonian (Eq.(\ref{Heff})), the energy dispersion relation $E(\vec{k})$ of lowest two bands can be gotten:
\begin{eqnarray}
\label{E-k}
E_{\pm}(\vec{k})=\mathcal{H}_{eff,0} \pm \sqrt{\mathcal{H}^2_{eff,x}+\mathcal{H}^2_{eff,y}+\mathcal{H}^2_{eff,z}}
\end{eqnarray}

Here we set $\delta=3/2$ to eliminate the constant term in the bracket of $\mathcal{H}_{eff,z}$, which makes this system better reflect the effects of shaking. Then, the energy disperion relation in Eq.(\ref{E-k}) mainly depends on parameter $\beta=\Delta/\hbar\omega$, and hopping coefficients $t_0$ and $t_1$ also depend on $\Delta$, (see Appendix A). And further discussion about different values of $\delta$ will be shown in Section VI. 

\section{The transformation from Dirac semimetal to nodal-line semimetal}
The value of $\mathcal{J}_{n}(\beta)$ will change as one of the shaking factor $\beta$ changes, as shown in Fig.\ref{fig:fig2}(a), where the different value of $\mathcal{J}_0$ and $\mathcal{J}_1$ corresponding to $\beta$. With $\mathcal{J}_1(\beta)=0$, $\mathcal{H}_{eff,z}$ is zero, but $\mathcal{H}_{eff,y}$ have a finite quantity. $E_{+}(\vec{k})$ and $E_{-}(\vec{k})$ equals to each other at six Dirac points, which performs as Dirac semimetal, as shown in Fig.\ref{fig:fig2}(b1). It is worth mentioning that although shaking breaks the inversion symmetry, the gap between the upper and lower band is still closed at Dirac point. Fig.\ref{fig:fig2}(b2) is the sectional view of as Fig.\ref{fig:fig2}(b1). In the figure, the blue line means the upper band and the dashed red line represents the lower band, which touches the former at the edge of the Brillouin zone.

In another situation for $\mathcal{J}_0(\beta)=0$, $\mathcal{H}_{eff,x}$ and $\mathcal{H}_{eff,y}$ is zero, but $\mathcal{H}_{eff,z}$ is nonzero. $E_{+}(\vec{k})$ and $E_{-}(\vec{k})$ equals to each other at a ring $k_x^2+k_y^2=(\frac{4 \pi}{9})^2/a^2$. This energy dispersion relation performs as nodal-line semimetal, as shown in Fig.\ref{fig:fig2} (c1). The upper and lower band touch inside the first Brillouin zone, which differs from the case in Dirac semimetal where the touchpoint is at the vertices of the first Brillouin zone. The most noteworthy feature of nodal-line semimetal is that nodes of two bands are continuous and form a so-called nodal-line ring, wherein our model locates at $k^2_x+k^2_y=(\frac{4 \pi}{9})^2/a^2$. Fig.\ref{fig:fig2}(c2) is sectional view of Fig.\ref{fig:fig2}(c1). The two bands touch at the nodal-line ring, and at the edge of the Brillouin zone, the gap between two bands is open.

\begin{figure}
	\includegraphics[width=0.5\textwidth]{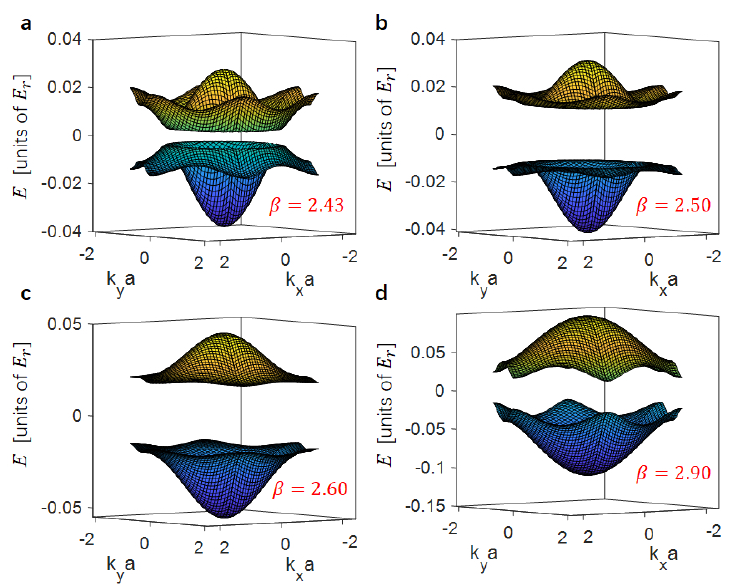}
	\caption{\textbf{The transformation of band structure as $\beta$ changes.} The band structure for the intermediate states of Dirac and node-line semimetal, at $\beta=2.43,2.50,2.60,2.90$ for (a)-(d), respectively, which show the intermediate states of two modes, which correspond to the four lines in Fig.\ref{fig:fig2}(a). Other parameters are chosen to be the same as Fig.\ref{fig:fig2}(b) and Fig.\ref{fig:fig2}(c).
	}\label{fig:fig4}
\end{figure}

When $\beta$ changes from $\mathcal{J}_0(\beta)=0$ to $\mathcal{J}_1(\beta)=0$, the system transforms from Dirac semimetal to nodal-line semimetal. Fig.\ref{fig:fig4} shows the band structure for the other value between $\mathcal{J}_0(\beta)=0$ and $\mathcal{J}_1(\beta)=0$ corresponding to the four vertical dot dashed red lines and two colorful curves in Fig.\ref{fig:fig2}(a). When the energy spectrum gradually varies from nodal-line semimetal to Dirac semimetal, the touching ring gradually opens as the Dirac points gradually close, as shown in Fig.\ref{fig:fig4}(a-d). In the experiment, $\beta$ can be changed continuously by fixing the amplitude of the lattice depth shaking and adjusting the rotating frequency continuously to observe the transformation from Dirac semimetal to nodal-line semimetal.

In Fig.\ref{fig:fig2} and Fig.\ref{fig:fig4}, the parameters are chosen from the experimental parameters of $^{87}Rb$ atom: $V_0=10.0 E_r$, $t_0=0.1538 E_r$, $t_1=0.0118 E_r$, $t_{NNN}=7.689 \times 10^{-4} E_r$, and $\Delta=0.50E_r$, where $E_r$ means atomic recoil energy of electron in our honeycomb lattice.

The transformation from Dirac semimetal to nodal-line semimetal can be explained by symmetry as follows. The impact of amplitude shaking is reflected in Hamiltonian, so we first study the Hamiltonian $\hat{H'}$. Through Jacobi-Anger expansion (ignoring terms unrelated to $\beta$), the Hamiltonian $\hat{H'}$ can be written as the following form:
\begin{eqnarray}
\label{H3}
\hat{H}'&&=-\sum_{\langle i,j\rangle ~and ~i \not = j }t_{ij} e^{iz_{ij}\sin\omega t} c_i^\dag c_j \\
&&=-\sum_{\langle ij\rangle} \mathcal{J}_0(\beta) t_{NN} c_i^\dag c_j 
-(e^{i\omega t}+e^{-i\omega t}) \nn \\
&&\cdot \sum_{\langle ij\rangle} \mathcal{J}_1(\beta) t_{NN} c_i^\dag c_j -{\rm o}(t_{NN} c_i^\dag c_j) \nn
\end{eqnarray}
Where $\sum_{\langle ij\rangle}$ means the sum of nearest-neighbor lattice nodes, and $o(t_{NN} c_i^\dag c_j)$ represents the higher order infinitesimal term considering the pairs of sites with farther distance between them than the next-nearest neighbor term comparing to $t_{NN} c_i^\dag c_j$. So the system mainly is composed of two modes, the weights of which are $\mathcal{J}_0(\beta)$ and $\mathcal{J}_1(\beta)$. 

\begin{figure}
	\includegraphics[width=0.5\textwidth]{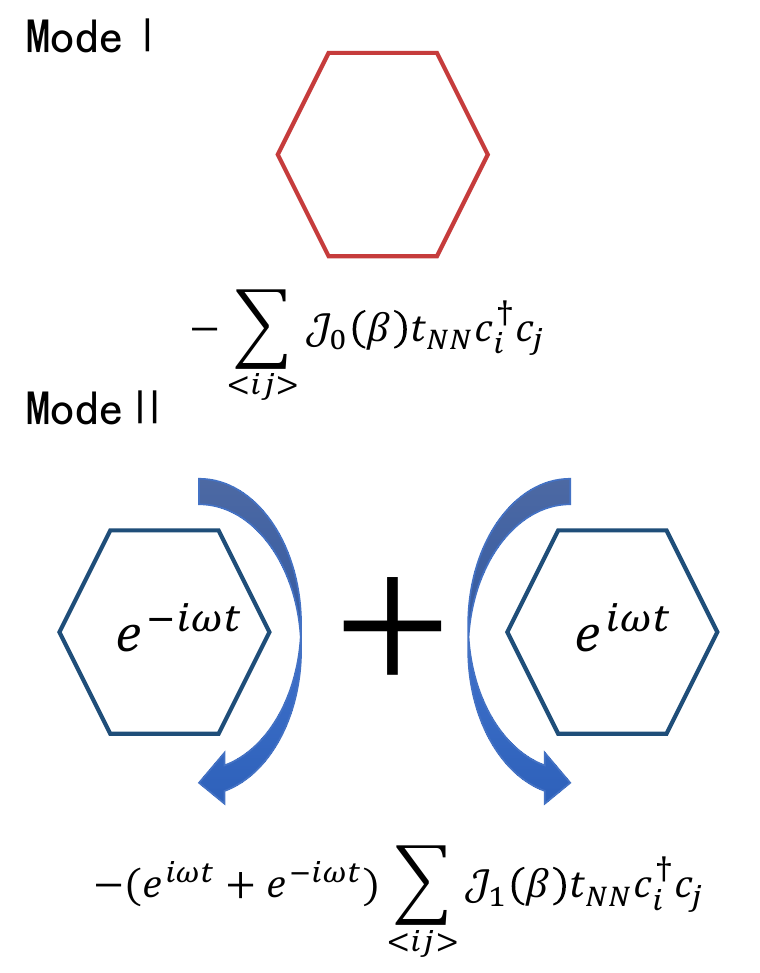}
	\caption{\textbf{Two shaking modes of the system.} The result of Fourier transform of Hamiltonian $\hat{H'}$ reflects two shaking modes of the system, where Mode \uppercase\expandafter{\romannumeral1} is a bipartite hexagonal lattice mode and Mode \uppercase\expandafter{\romannumeral2} is a nontrivial time-reversal-symmetry-unstable mode, which corresponds to nodal-line semimetal.
	}\label{fig:fig3}
\end{figure}

As shown in Fig.\ref{fig:fig3}, the expanding term of Hamiltonian $\hat{H'}$ can be divided into two parts. The first term corresponds to Mode \uppercase\expandafter{\romannumeral1}, which is a bipartite hexagonal lattice Mode.
When $\mathcal{J}_1(\beta)=0$, only mode \uppercase\expandafter{\romannumeral1} exists in the system, so the system performs like an ordinary Dirac semimetal. In the situation, Dirac points are closed due to the spatial inversion symmetric potential energy in Mode \uppercase\expandafter{\romannumeral1}.

The second term of Hamiltonian $\hat{H'}$ corresponds to Mode \uppercase\expandafter{\romannumeral2}, which consists of two symmetrical rotating lattices. The Mode \uppercase\expandafter{\romannumeral2} is the same as Mode \uppercase\expandafter{\romannumeral1}, both of which have time-reversal symmetry, but different from Mode \uppercase\expandafter{\romannumeral1}, the time inversion symmetry of Mode \uppercase\expandafter{\romannumeral2} is unstable. In our system, the phase difference is $\pi$, so the difference of well depth can be written as a cosine function, and the amplitude of two sub-modes are equivalent to each other. If we change the phase difference, the form of well depth difference between point A and B change consequently, which causes the two sub-modes of Mode \uppercase\expandafter{\romannumeral2} to be asymmetric and leads to the broken of time-reversal symmetry. When $\mathcal{J}_0(\beta)=0$, Mode \uppercase\expandafter{\romannumeral1} will vanish while Mode \uppercase\expandafter{\romannumeral2} remains existing in the system, so its valence and conducting band touch at a ring, performing as a nodal-line semimetal. When $\beta$ is between the zeros of $\mathcal{J}_0$ to $\mathcal{J}_1$, the system is in the intermediate state of two modes.

\section{Berry curvature and Berry phase}
Considering that this is a two-dimensional system, in this section, we calculate the Berry curvature and Berry phase rather than winding number to provide a measurable quantity in the experiment during the transformation from Dirac semimetal to nodal-line semimetal.

From Eq.(\ref{Heff}), we can define $\vec{h}$ as:
\begin{eqnarray}
\label{h}
\vec{h}=\mathcal{H}_{eff,x}\cdot\hat{e}_x+\mathcal{H}_{eff,y}\cdot\hat{e}_y+\mathcal{H}_{eff,z}\cdot\hat{e}_z
\end{eqnarray}
where $\hat{e}_x,\hat{e}_y,\hat{e}_z$ are the basis vectors at $x$, $y$, $z$ direction of momentum space. The role that $\vec{h}$ plays is similar to a magnetic field coupling with the vector of Pauli matrices $(\hat{\sigma}_x,\hat{\sigma}_y,\hat{\sigma}_z)$. By solving the eigenequations of effective Hamiltonian, we get 

\begin{eqnarray}
\label{eigenstates+}
\psi_+=\dfrac{1}{\sqrt{2h(h+h_3)}}\begin{pmatrix}
h_3+h \\ h_1-\mathbbm{i}h_2
\end{pmatrix} 
\end{eqnarray}
\begin{eqnarray}
\label{eigenstates-}
\psi_-=\dfrac{1}{\sqrt{2h(h-h_3)}}\begin{pmatrix}
h_3-h \\ h_1-\mathbbm{i}h_2
\end{pmatrix} 
\end{eqnarray}
where $\vec{h}=h_1\hat{e}_x+h_2\hat{e}_y+h_3\hat{e}_z$, $h=\sqrt{h_1^2+h_2^2+h_3^2}$.

Two components $\mathcal{A}_i \,(i=1,2)$ of Berry connection $\vec{\mathcal{A}}$ of the lowest band can be calculated from $\psi_-$ as
\begin{eqnarray}
	\label{Bconnection}
	\mathcal{A}_i(\vec{k})=\mathbbm{i}\langle \psi_-|\partial_{k_i}|\psi_-\rangle=-\dfrac{h_2\partial_{k_i}h_1-h_1\partial_{k_i}h_2}{2h(h-h_3)}
\end{eqnarray}

Then the Berry curvature of the lowest band can be calculated as
\begin{eqnarray}
\label{berry curvature}
\vec{\Omega}(\vec{k})=-\nabla\times\vec{\mathcal{A}}(\vec{k})
\end{eqnarray}

The $z$ component of the Berry curvature is
\begin{eqnarray}
\label{berry curvature z}
\Omega_3=-\dfrac{\partial \mathcal{A}_2}{\partial k_1}+\dfrac{\partial \mathcal{A}_1}{\partial k_2}
\end{eqnarray}

Hence the Berry phase can be calculated from Berry curvature:
\begin{eqnarray}
\label{berry phase}
\gamma=\int _S {\rm d}\vec{S} \cdot  \vec{\Omega}(\vec{k})=\int _S {\rm d}S\, \Omega_3(\vec{k})
\end{eqnarray}

\begin{figure}
	\includegraphics[width=0.5\textwidth]{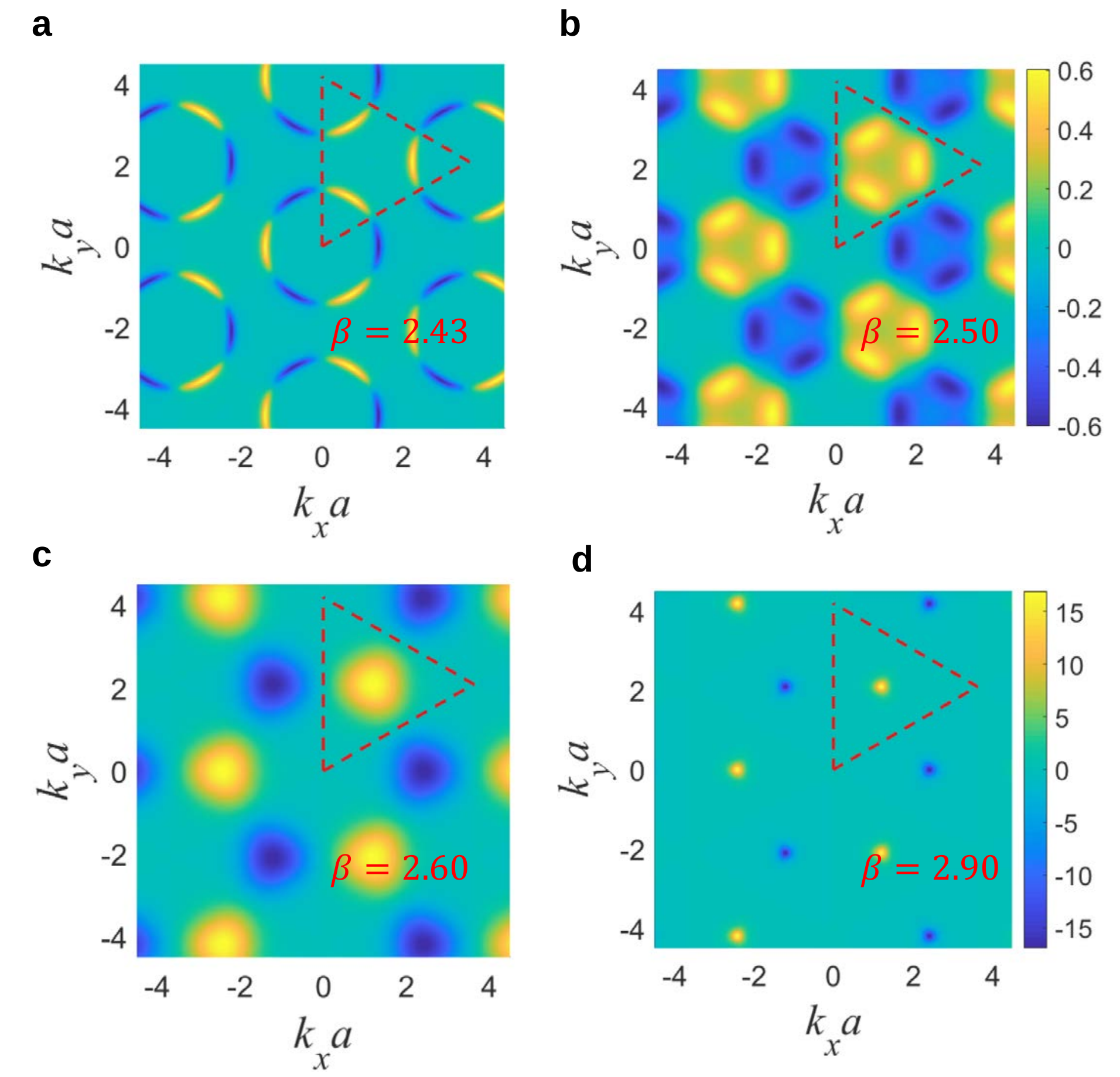}
	\caption{\textbf{The change of Berry curvature during the transformation process.} (a-d) Berry curvature in the reciprocal space in the process of the system transforming from nodal-line semimetal to Dirac semimetal.  $\beta=2.43,2.50,2.60,2.90$, which is correspond to the four lines in Fig.\ref{fig:fig2}(a), respectively. The dashed equilateral triangle denotes the integral region $S$ of the Berry phase, which goes around one vertex of the first Brillouin Zone.
	}\label{fig:fig5}
\end{figure}


Here $S$ denotes the area in the reciprocal space. Fig.\ref{fig:fig5} shows the Berry curvature in the transformation from nodal-line semimetal to Dirac semimetal. In the figure, the values of $\beta$ are selected according to the intersections of four vertical dot dashed red lines and two colorful curves in Fig.\ref{fig:fig2} (a). When $\beta$ is near 2.4048 (one of the zeros of $\mathcal{J}_0$), the system is nearly a nodal-line semimetal, and Berry curvature of which is nontrivial at the nodal-line ring. The distribution of Berry curvature can naturally divide into six parts, where the numerical value of adjacent parts is of opposite sign, and the sum of six parts equals to zero, which is protected by time-reversal symmetry.
With the system transforming to Dirac semimetal, while $\beta$ increases from 2.4048 to 3.8317 (the nearest zero of $\mathcal{J}_1$), those positive and negative parts of Berry curvature respectively become closer together. Finally, those parts converge on six Dirac points and keep shrinking to approach the distribution in the case of Dirac semimetal. Because the shrinking of Berry curvature into one point makes it difficult to distinguish in the figure, here we only show the result up to $\beta=2.90$ in Fig.\ref{fig:fig5}. 

\begin{figure}
	\includegraphics[width=0.5\textwidth]{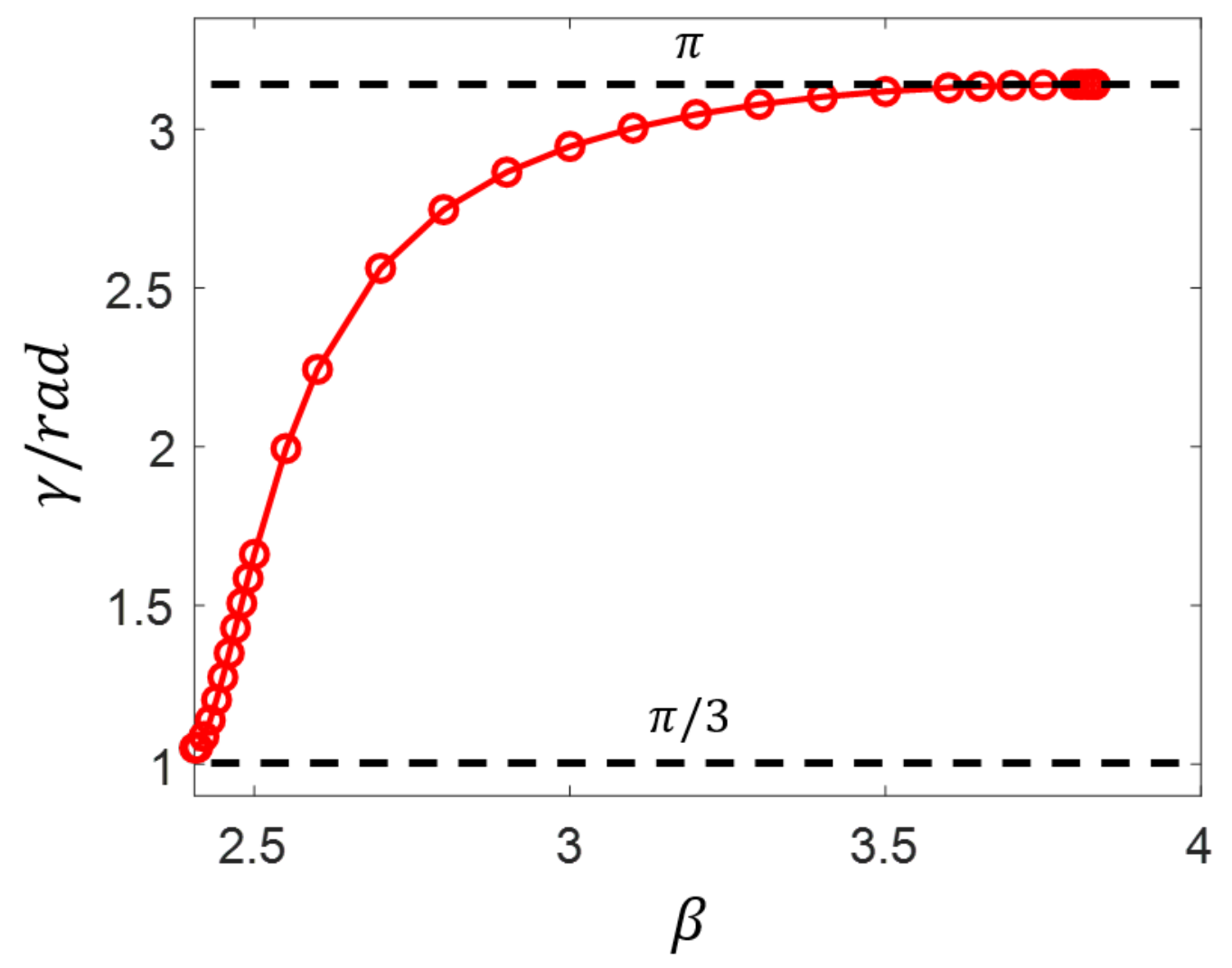}
	\caption{\textbf{Berry phase around a Dirac point during the transformation.} Here we set $\delta=3/2$. In the figure, the red circles are simulated data points, and the red line is fitted curve. The two black dotted lines mark the start and end Berry phase during the transformation. When the system is a Dirac semimetal, the Berry phase is equal to $\pi$. And when the system transforms to nodal-line semimetal, the Berry phase changes continuously to $\pi/3$.
	}\label{fig:fig6}
\end{figure}

The red dashed equilateral triangle which connects the center of one site and two adjacent sites in Fig.\ref{fig:fig5} shows the path around one Dirac point, and we study the change of the Berry phase along the path in the transformation. Generally the path to calculate berry phase should avoid the passing Berry-curvature singularities \cite{PhysRevB.84.205440}, but in our system, Berry-curvature singularities will change from Dirac points to nodal-line ring during the transformation. It is impossible to choose a path along a Dirac point disjointing with any singularities in a transformation. Hence we choose the equilateral triangle path which is along a Dirac point and passes high symmetry point of Brillouin zone.
Fig.\ref{fig:fig6} shows the result of the Berry phase as $\beta$ increases from 2.408 to 3.70 (between zeros of $\mathcal{J}_0$ 2.4048 to $\mathcal{J}_1$ 3.8317). When the system is nearly a nodal-line semimetal, the Berry phase becomes close to $\pi/3$, which corresponds to the situation in Fig.\ref{fig:fig5}(a). When the system is near Dirac semimetal, our Floquet system gets the same result as in bipartite hexagonal lattice, and the Berry phase around Dirac point comes close to $\pi$.  In the transformation procession, as is shown in Fig.\ref{fig:fig5}(a)-(d), the Berry phase around Dirac point increases continuously, with the Berry curvature shrinking into one point.

\begin{figure}
 	\includegraphics[width=0.5\textwidth]{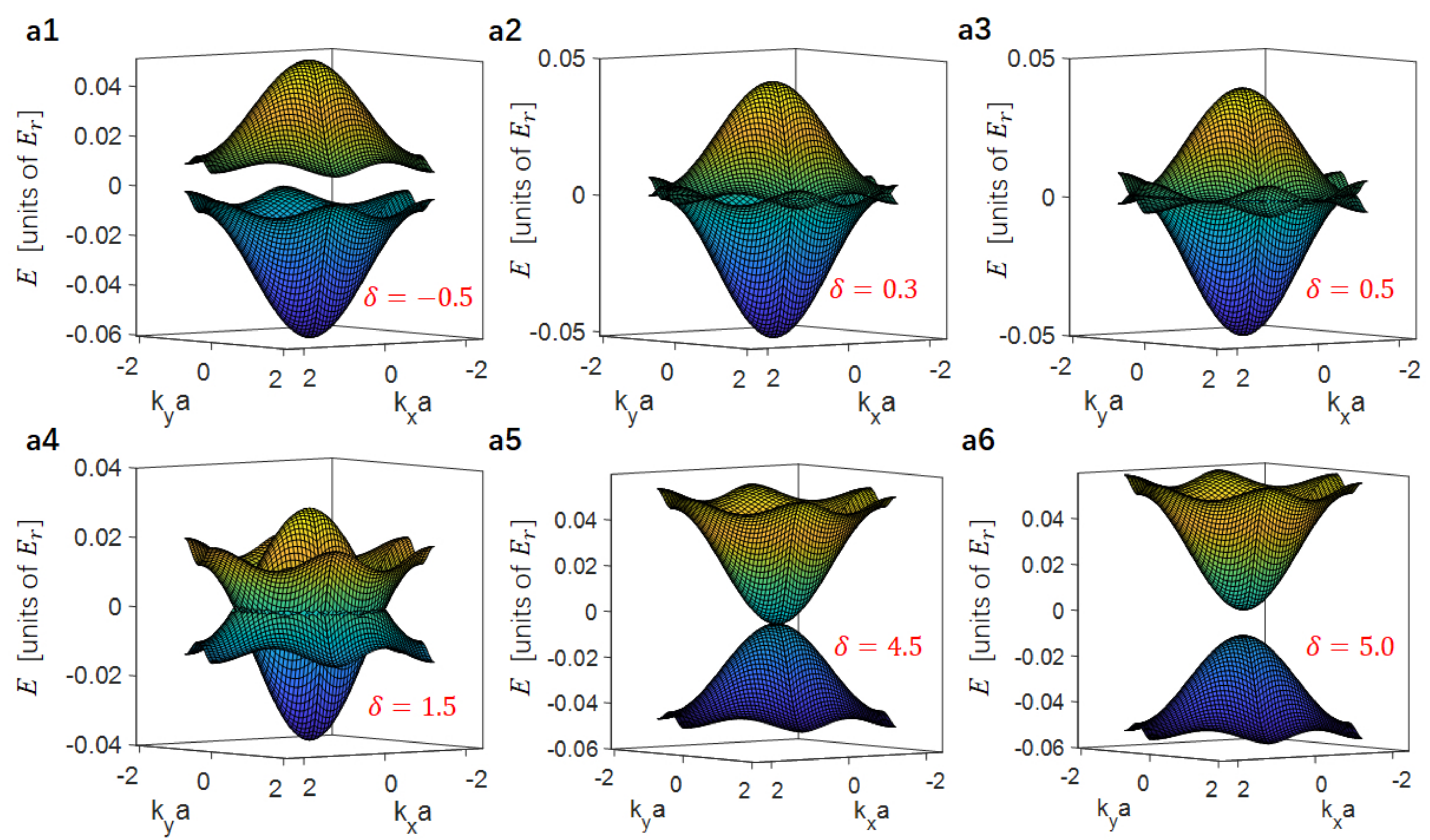}
 	
 	\includegraphics[width=0.5\textwidth]{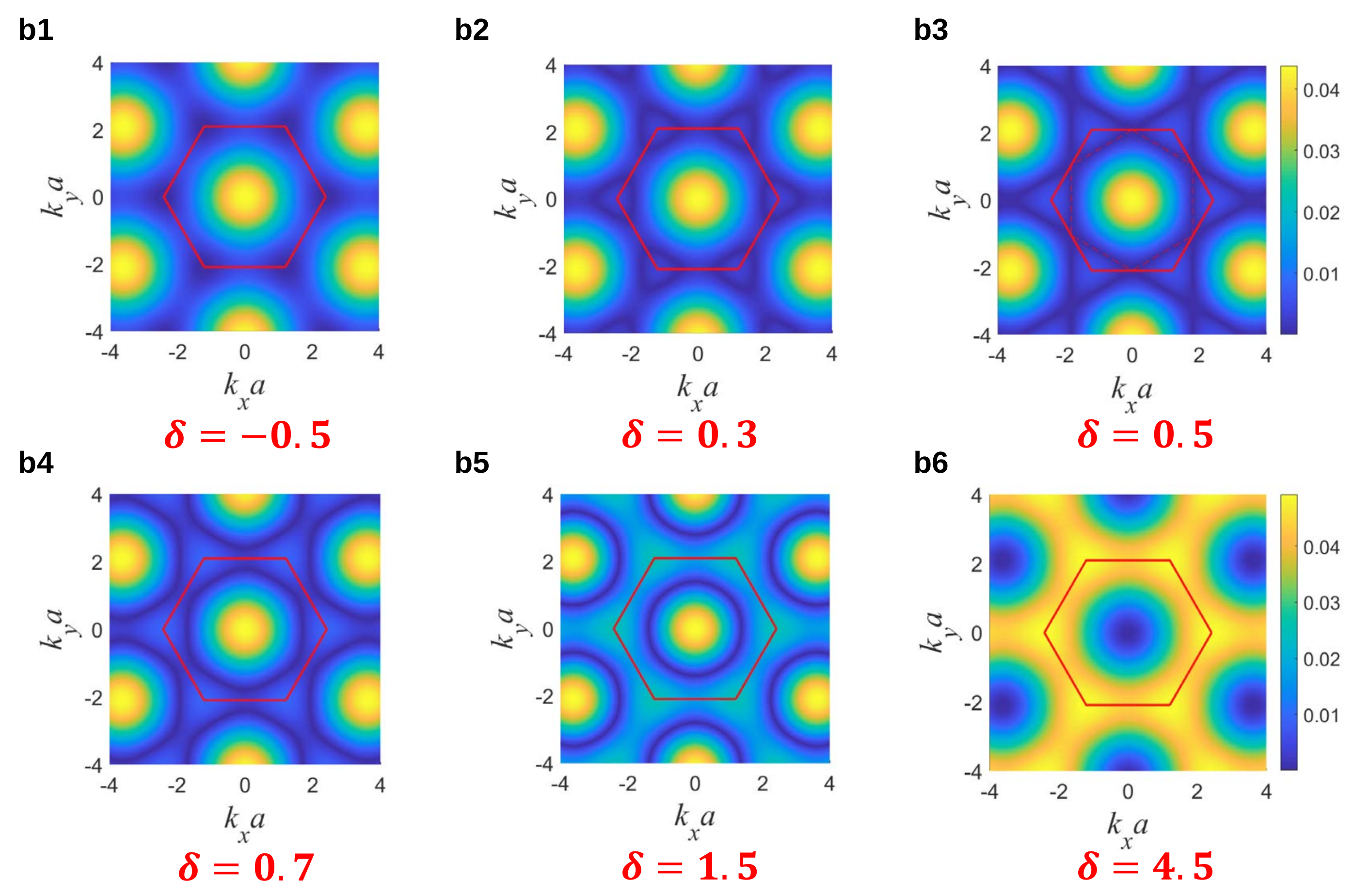}
 	\caption{\textbf{The band structure of nodal-line phase with $\delta $ changes.} (a1-a6) The front view of the band structure for the nodal-line phase at $\beta=2.4048$, where $\delta=-0.5,\,0.3,\,0.5,\,1.5,\,4.5,\,5.0$ from (a1) to (a6), respectively. (b1-b6) The top view of the band structure for the nodal-line phase at $\beta=2.4048$, where $\delta=-0.5,\,0.3,\,0.5,\,0.7,\,1.5,\,4.5.0$ from (b1) to (b6), respectively.  The red solid hexagon denotes the boundary of the first Brillouin zone.
 	}\label{fig:fig8}
 \end{figure}

Above, we focus $\delta=1.5$ and adjust $\beta$ to observe the transformation from Dirac semimetal to Nodal-line semimetal. Next, we will set $\beta=2.4048$ and study the change of nodal-line phase while the value of $\delta$ changes. The band structure of the nodal-line phase with different $\delta$ is shown in Fig.\ref{fig:fig8}(a). In the figure, through changing the well difference $\delta$, different band structure of nodal-line phase is gotten. When $\delta<0$, the upper and lower bands don't intersect with each other; when $\delta\in(0,0.5)$, the nodal-line appears in the first Brillouin zone, they are six discrete curves with $C_6$ symmetry; when $\delta\in(0.5,1.5)$, the nodal-line becomes a closed unrounded loop around the center of the first Brillouin zone; when $\delta\in(1.5,4.5)$, the nodal-line is a circle, the radius of which decreases as $\delta$ increases. When $\delta=4.5$, there is a single node at the center of the first Brillouin zone, and two bands will be gapped for $\delta>4.5$.

 The change of the Berry phase around the closed path in Fig.\ref{fig:fig5} with the change of $\delta$ and symmetry analysis are studied for fixed $\beta=2.4048$ in nodal-line phases(See Appendix D).

 In our system, the Chern number is not a good measurable quantity. Because, in our system, the Chern number is always zero. The integral of Berry curvature along the edge of the first Brillouin zone gives the Chern number, due to the protection of time-reversal symmetry, the integral is always zero.
 
 Recently many experiments implemented with ultracold atoms in an optical lattice system have focused on studying the Berry curvature or other topological invariant \cite{Aidelsburger2015,Zhang_2016,Asteria2019}, and some researchers among them have developed mature technology to map Berry curvature \cite{PhysRevA.85.033620}. So our result of Berry curvature and Berry phase can be verified in the future experiments by existing technology.

\section{Conclusions}
In summary, we propose a feasible scheme to simulate nodal-line semimetal with ultracold atoms in an amplitude-shaken optical lattice. We derive the effective Hamiltonian of the Floquet system, and by calculating the band structure, the transformation from Dirac semimetal to nodal-line semimetal is observed. When the shaking factor $\beta$, which is determined by the shaking amplitude and frequency, is at zeros of the first-order Bessel function of the first kind, the band structure performs as Dirac semimetal, and when shaking factor is at the zeros of the zeroth-order Bessel function, the band structure performs as a nodal-line semimetal. Through Fourier transform, we divide the shaking into two modes, which explain the above transformation. The change of Berry curvature and Berry phase during the transformation process shows the topological characteristics of our system. However, if we set the phase difference of amplitude oscillation of sites A and B to not be $\pi$ exactly, the time-reversal symmetry of the system will be broken, which may lead to the research of new topological semimetals.

\section{Acknowledgement}
We thank Xiaopeng Li, Wenjun Zhang, Yuan Zhan  for helpful discussion. This work is supported by the National Basic Research Program of  China (Grant No. 2016YFA0301501) and the National Natural Science Foundation of China (Grants No. 61727819, No. 11934002, No. 91736208, and No. 11920101004).

\appendix
\addcontentsline{toc}{section}{Appendices}\markboth{APPENDICES}{}
\begin{subappendices}
\section{hopping coefficient}
We estimate the nearest-neighbor hopping coefficient in our Floquet system, taken $^{87}Rb$ atom for instance. For a hexagonal optical lattice with different well depth at points A and B, the overlapping integral of Wannier function helps us to calculate the nearest-neighbor hopping coefficient $t_{NN}$, which renders result consistent with Equation~(\ref{tnn}).

\begin{figure}
	\includegraphics[width=0.5\textwidth]{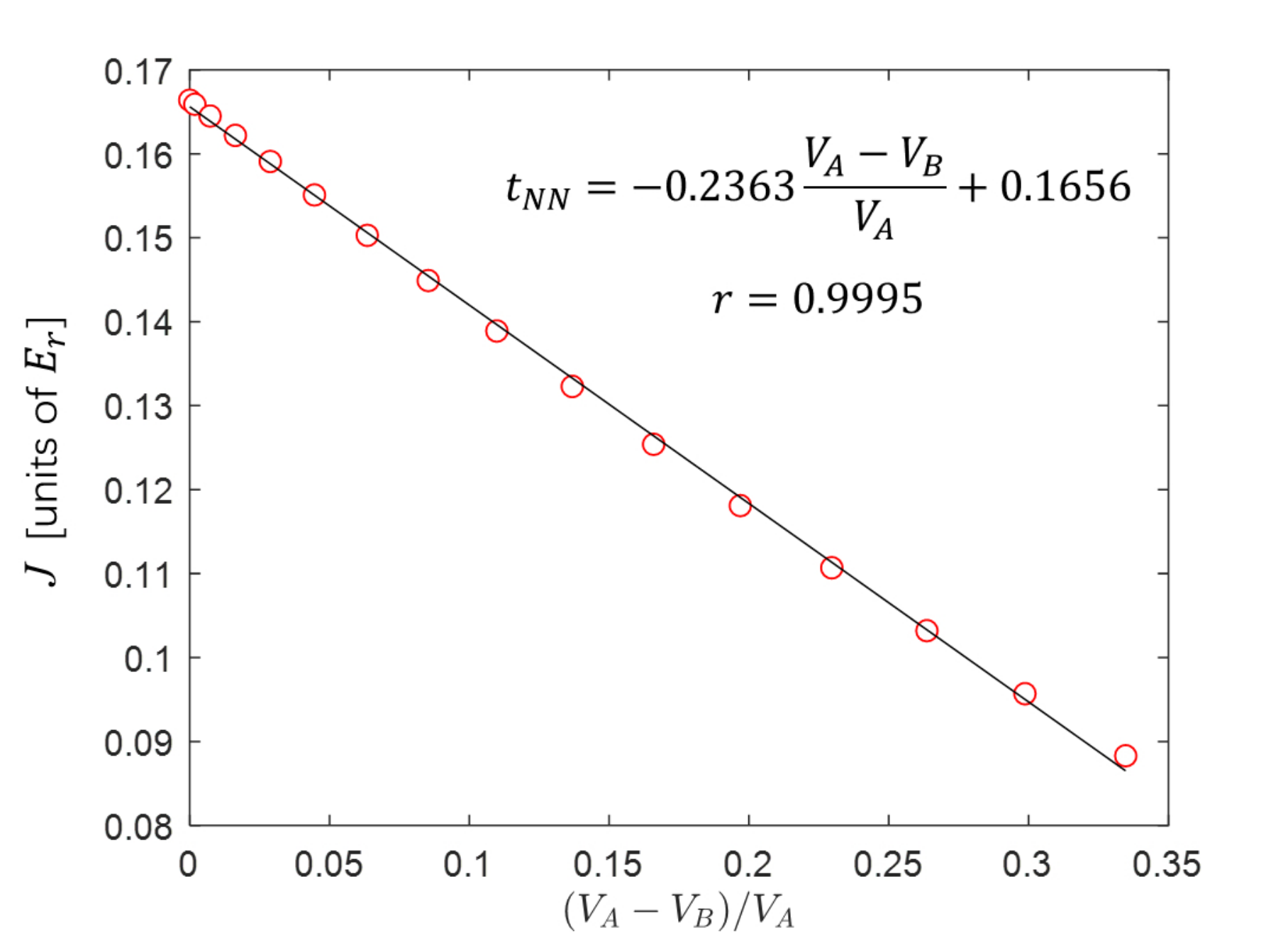}
	\caption{\textbf{The theoretical calculation of the nearest neighbor hopping coefficient $J$ as difference of well depth changes.} The nearest neighbor hopping coefficient in a hexagonal optical lattice nearly changes linearly as the well depth difference between the lattice depth of point A and B changes. The correlation coefficient $r$ is 0.9995, which proves the good linearity. In the figure, the red circle is the calculated data and the line is the fitting line. The well depth at point $A$ is about $10 E_r$, where $E_r$ means the atomic recoil energy in our honeycomb lattice.
	}\label{fig:S1}
\end{figure}

Fig.\ref{fig:S1} shows the change of nearest-neighbor hopping coefficient as the difference of well depth $(V_A-V_B)/V_A$ changes. As a result, the nearest-neighbor hopping coefficient $t_{NN}$ equals one constant $t_0$ adding one term which is nearly proportional to the lattice depth difference between points A and B in a large range (Correlation coefficient is $0.9995$). In the process of changing the lattice depth, the change of well depth at point A $V_A$ is less than $20\%$, and the difference of well depth $(V_A-V_B)/V_A$ is about $35\%$. The linearity will be better if the process happens in a smaller range. So the static process can be used to estimate our dynamic perturbation process. In our system, the lattice depth difference between points A and B changes over time as a cosine function, which leads to the conclusion that the nearest-neighbor hopping coefficient also changes as a cosine function, so we use $t_{NN}=t_0+t_1\cos(\omega t)$ to estimate it in the calculation. Using the fitting result, we can calculate that

\begin{eqnarray}
\label{t1}
t_1=0.2363\dfrac{\Delta}{V_A}E_r
\end{eqnarray}

\begin{eqnarray}
\label{t0}
t_0=0.1656E_r-0.2363\dfrac{\Delta}{V_A}E_r
\end{eqnarray}
where $E_r$ means the atomic recoil energy of electron in our honeycomb lattice. By substituting $\Delta=0.5E_r$ and $V_A=10E_r$, we get
\begin{eqnarray}
	\label{t0t1num}
	t_0=0.1538E_r,\quad t_1=0.0118E_r
\end{eqnarray}

As for the next-nearest-neighbor hopping coefficient, because the lattice depth of the next neighbor point is always $0$, we use a constant to replace the next neighbor hopping coefficient
\begin{eqnarray}
\label{tNNNnum}
t_{NNN}=0.0050E_r
\end{eqnarray}

In addition, according to Eq.(\ref{H eff}), the high-frequency expansion requires shaking frequency $\omega$ to be far greater than $H_{f0}/\hbar$. In our system, it requires that difference of the lattice depth $\Delta$ is larger than 0.15 $V_A$, which meets the linear range.

\section{Unitary transformation of Hamiltonian}
The unitary operator $\hat{U}$ can be obtained by substituting Equation~(\ref{depth of A and B}) into Equation~(\ref{U}):
\begin{eqnarray}
\label{U1}
\hat{U}(t)&&=\exp\left[\dfrac{\mathbbm{i}}{\hbar} \sum_{i}\int_{0}^{t}d\tau \cdot \chi(i)\frac{\Delta}{2} \cos(\omega \tau)c_i^\dag c_i\right] \nn \\
&&=\exp\left[\dfrac{\mathbbm{i}}{\hbar} \sum_{i} \chi(i)\frac{\Delta}{2\omega} \sin(\omega \tau)c_i^\dag c_i\right] 
\end{eqnarray}

Then we get
\begin{eqnarray}
\label{U on partial}
\hat{U}\left(-\mathbbm{i} \hbar\dfrac{\partial}{\partial t}\right)\hat{U}^\dag&&=-\mathbbm{i} \hbar\dfrac{\partial}{\partial t}-\sum_{i}\frac{\Delta}{2} \cos(\omega t)\chi(i)c_i^\dag c_i  \nn\\
&&
\end{eqnarray}

Define $\hat{U}_k'(t)$ as
\begin{eqnarray}
\label{U_k}
\hat{U}_k'(t)=\exp\left[\left(\dfrac{\mathbbm{i}}{\hbar}\dfrac{\Delta}{2\omega}\sin\omega t \chi(k)\right)c_k^\dag c_k\right]
\end{eqnarray}

According to Baker-Campbell-Hausdorff formula,
\begin{eqnarray}
\label{Baker}
e^{\hat{A}}\hat{B}e^{-\hat{A}}=\hat{B}+[\hat{A},\hat{B}]+\dfrac{1}{2!}[\hat{A},[\hat{A},\hat{B}]]+\dots
\end{eqnarray}

We can get
\begin{eqnarray}
\label{U on cicj}
\hat{U}(-t_{ij}c_i^\dagger c_j)\hat{U}^\dag&&=\sum_k \hat{U}_k'(-t_{ij}c_i^\dagger c_j)\sum_k \hat{U}_k'^\dagger \nn \\
&&=-t_{ij}e^{\frac{\mathbbm{i}}{\hbar}\frac{\Delta}{2\omega}\sin \omega t[\chi(i)-\chi(j)]}c_i^\dagger c_j
\end{eqnarray}

The transformed Hamiltonian was finally obtained
\begin{eqnarray}
\label{H'}
\label{U1}
\hat{H'}&&=\hat{U}\left(\hat{H}(t)-\mathbbm{i} \hbar\dfrac{\partial}{\partial t}\right)\hat{U}^\dag-\left(-\mathbbm{i} \hbar\dfrac{\partial}{\partial t}\right)  \\
&&=\sum_i V_i(t)c_i^\dagger c_j-\sum_{i \not = j }t_{ij} e^{iz_{ij}\sin\omega t} c_i^\dag c_j  \nn \\
&&-\mathbbm{i} \hbar\dfrac{\partial}{\partial t}-\sum_{i}\frac{\Delta}{2} \cos(\omega t)\chi(i)c_i^\dag c_i+\mathbbm{i} \hbar\dfrac{\partial}{\partial t}  \nn \\
&&=-\dfrac{d}{2}\sum_i (a_i^\dag a_i-b_i^\dag b_i)-\sum_{i \not = j }t_{ij} e^{\mathbbm{i}z_{ij}\sin\omega t} c_i^\dag c_j \nn
\end{eqnarray}
where the first term in Equation~(\ref{H 0}) was neglected, since the zero point of energy has been set at $V_0$, and $z_{ij}\equiv\frac{\Delta}{2 \omega \hbar}[\chi(i)-\chi(j)]$. 

\section{Fourier expansion and derivation of effective Hamiltonian}
By a Jacobi-Anger expansion, 
$e^{\mathbbm{i}z\sin\theta}=\sum_{n=-\infty}^{+\infty}\mathcal{J}_n(z)e^{\mathbbm{i}n\theta}$, we expand $\hat{H'}$ in Eq.(\ref{H tr}),  taking both nearest-neighbor and next-nearest-neighbor hopping into consideration, then
\begin{eqnarray}
\label{H' expand}
\hat{H'}(t)
&&=-\dfrac{d}{2}\sum_i (a_i^\dagger a_i-b_i^\dagger b_i) \nn \\
&&-\sum_{n=-\infty}^{+\infty}e^{\mathbbm{i}n\omega t}\left(\sum_{\langle ij\rangle}\left(t_0\mathcal{J}_n(z_{ij})\right.\right. \nn \\
&&
\left.+\dfrac{t_1}{2}[\mathcal{J}_{n-1}(z_{ij})+\mathcal{J}_{n+1}(z_{ij})]\right)c_i^\dagger c_j \nn \\
&&+\left. \sum_{\langle\langle ij\rangle\rangle}\mathcal{J}_n(0)t_{NNN}c_i^\dagger c_j\right) 
\end{eqnarray}
where $\sum_{\langle ij\rangle}$ denotes the summation over the nearest-neighbor nodes, and $\sum_{\langle\langle ij\rangle\rangle}$ denotes the summation over the next-nearest-neighbor nodes. Then we can obtain the Fourier expansion coefficients of Hamiltonian for $n>0$
\begin{eqnarray}
\label{Hn}
H_n&&=-\sum_{\langle ij\rangle}\left(t_0+t_1\dfrac{n}{z_{ij}}\right)\mathcal{J}_n(z_{ij})c_i^\dagger c_j \nn \\
&&-\sum_{\langle\langle ij\rangle\rangle}\mathcal{J}_n(0)t_{NNN}c_i^\dagger c_j
\end{eqnarray}

From the Floquet theory, we can derive the formula to calculate effective Hamiltonian\cite{PhysRevX.4.031027,PhysRevX.5.029902}
\begin{eqnarray}
\label{heff}
H_{eff}=H_{eff}^0+\dfrac{1}{\hbar\omega}H_{eff}^1=H_{f0}+\sum_{n=1}^{\infty} \frac{[H_{n},H_{-n}]}{n \hbar \omega} \nn\\
&&
\end{eqnarray}
where higher-order terms which consider the pairs of sites with the farther distance between them than the next-nearest neighbor term have been neglected due to high-frequency approximation. The calculation of substituting Eq.(\ref{Hn}) into Eq.(\ref{heff}) is shown below. 

For $n=0$ term,

\begin{eqnarray}
\label{H f0}
H_{f0}&&=-\sum_{\langle ij\rangle}t_0\mathcal{J}_n(z_{ij})c_i^\dagger c_j-\sum_{\langle\langle ij\rangle\rangle}t_{NNN}c_i^\dagger c_j \nn \\
&&-\dfrac{d}{2}\sum_i (a_i^\dagger a_i-b_i^\dagger b_i) \nn \\
&&=H_{0,1}+H_{0,2}+H_{0,3}
\end{eqnarray}
where $H_{0,1}$ denotes the summation over the nearest-neighbor nodes, and $H_{0,2}$ denotes the summation over the next-nearest-neighbor nodes. For $H_{0,1}$, because $\mathcal{J}_0(-z_{ij})=\mathcal{J}_0(z_{ij})$, we can get

\begin{eqnarray}
\label{H01'}
H_{0,1}&&=-t_0\mathcal{J}_n(\beta)\sum_{i}(a_{\vec{r_i}}^\dagger b_{\vec{r_i}+\vec{u_1}} \nn \\
&&+a_{\vec{r_i}}^\dagger b_{\vec{r_i}+\vec{u_2}}+a_{\vec{r_i}}^\dagger b_{\vec{r_i}+\vec{u_3}})+h.c.
\end{eqnarray}
where $\beta\equiv\frac{\Delta}{\hbar\omega}$, $a_i$, $b_j$ denote the annihilation operators of lattice site $A_i$, $B_j$. The first term describes the hopping from $B_j$ to $A_i$, while the second term is the Hermitian conjugate of the former one, describes the hopping from $A_i$ to $B_j$. The $\vec{r_i}$ is the lattice vector for bipartite hexagonal lattice, and the  $\vec{u_i}$ is the nearest-neighbor vector.

Considering the creation and annihilation operators  as periodic functions in real space, we can take the Fourier transformation to get the corresponding creation and annihilation operators in momentum space $a_i=\dfrac{1}{\sqrt{N}}\sum_k a_k e^{-\mathbbm{i}\vec{k}\cdot\vec{r}} $ and $b_i=\dfrac{1}{\sqrt{N}}\sum_k b_k e^{-\mathbbm{i}\vec{k}\cdot\vec{r}} $, where $N$ is the number of lattice sites. By substituting these equations into the effective Hamiltonian in Eq.(\ref{H01'}), using $\delta_{\vec{k}\vec{k'}}=\frac{1}{N}\sum_i e^{\mathbbm{i}(\vec{k}-\vec{k'})\cdot\vec{r_i}}$, we can obtain that
\begin{eqnarray}
\label{H01''}
H_{0,1}&&=-t_0\mathcal{J}_n(\beta)\sum_{k}a_k^\dagger b_k \nn \\
&&(e^{-\mathbbm{i}\vec{k}\cdot\vec{u_1}}+e^{-\mathbbm{i}\vec{k}\cdot\vec{u_2}}+e^{-\mathbbm{i}\vec{k}\cdot\vec{u_3}})+h.c. \nn \\
&&=\sum_k 
\begin{pmatrix}
a_k^\dagger & b_k^\dagger
\end{pmatrix}
\begin{pmatrix}
0 & \mathcal{H}_{01}(k) 
\\ \mathcal{H}_{01}^\star(k) & 0
\end{pmatrix}
\begin{pmatrix}
a_k \\ b_k
\end{pmatrix}
\end{eqnarray}
where $\mathcal{H}_{01}(k)=-t_0\mathcal{J}_0(\beta)\sum_{j=1}^{3}e^{-\mathbbm{i}\vec{k}\cdot\vec{u_j}}$

For $H_{0,2}$, denote the lattice constant of bipartite hexagonal lattice as $\vec{v_j}$, as shown in Fig.\ref{fig:fig1}(a). Fourier transform and direct calculation gives that
\begin{eqnarray}
\label{H02}
H_{0,2}&&=-\sum_{\langle\langle ij\rangle\rangle}t_{NNN}c_i^\dagger c_j \nn \\
&&=-2t_{NNN}\sum_{j=1}^{3}\sum_k(a_k^\dagger a_{k}+b_k^\dagger b_{k})\cos(\vec{k}\cdot\vec{v_j}) \nn \\
&&=\sum_k 
\begin{pmatrix}
a_k^\dagger & b_k^\dagger
\end{pmatrix}
\begin{pmatrix}
\mathcal{H}_{02}(k) & 0 
\\ 0 & \mathcal{H}_{02}(k)
\end{pmatrix}
\begin{pmatrix}
a_k \\ b_k
\end{pmatrix}
\end{eqnarray}
where $\mathcal{H}_{02}=-2t_{NNN}\sum_{j=1}^{3}\cos(\vec{k}\cdot\vec{v_j})$.

As for $H_{0,3}$,

\begin{eqnarray}
\label{H03}
H_{03}
&&=
-\dfrac{d}{2}\sum_i (a_i^\dagger a_i-b_i^\dagger b_i)\nn \\
&&=
-\dfrac{d}{2}\sum_k (a_k^\dagger a_k-b_k^\dagger b_k)\nn \\
&&=\sum_k 
\begin{pmatrix}
a_k^\dagger & b_k^\dagger
\end{pmatrix}
\begin{pmatrix}
\mathcal{H}_{03} & 0 
\\ 0 & -\mathcal{H}_{03}
\end{pmatrix}
\begin{pmatrix}
a_k \\ b_k
\end{pmatrix} 
\end{eqnarray}
where$\mathcal{H}_{03}=-d/2$.

For $n>0$ terms, 
\begin{eqnarray}
\label{H0}
H_n
&&=-\sum_{\langle ij\rangle}\left(t_0\mathcal{J}_n(z_{ij})+\dfrac{t_1}{2}[\mathcal{J}_{n-1}(z_{ij})+\mathcal{J}_{n+1}(z_{ij})]\right)c_i^\dagger c_j \nn \\
\end{eqnarray}
According to Eq.(\ref{heff}),
\begin{eqnarray}
\label{H eff(1)}
H_{eff}^{(1)}&&=\sum_{n=1}^{\infty}\dfrac{[H_n,H_{-n}]}{n}\\
&&=\dfrac{4t_0t_1}{\beta}\mathcal{J}_1(\beta)^2\nn \\
&&\cdot\left\{3\sum_i[a_i^\dagger a_i-b_i^\dagger b_i]   +\sum_{\langle \langle ij\rangle\rangle}[a_i^\dagger a_j-b_i^\dagger b_j]\right\} \nn \\
&&=\sum_k 
\begin{pmatrix}
a_k^\dagger & b_k^\dagger
\end{pmatrix}
\begin{pmatrix}
\mathcal{H}_{1} & 0 
\\ 0 & -\mathcal{H}_{1}
\end{pmatrix}
\begin{pmatrix}
a_k \\ b_k
\end{pmatrix} \nn
\end{eqnarray}
which uses the commutation relationship $[c_k^\dagger c_l,c_p^\dagger c_q]=\delta_{lp}c_k^\dagger c_q-\delta_{kq}c_p^\dagger c_l$. And

\begin{eqnarray}
\label{H_1}
\mathcal{H}_{12}=8t_0t_1 \dfrac{\mathcal{J}_{1}(\beta)^2}{\beta} \left(\sum\limits_{j=1}^{3}\cos(\vec{k}\cdot\vec{v_j})+\dfrac{3}{2}\right)
\end{eqnarray}

Combining the results of Eq.(\ref{H01''}),~(\ref{H02}) and~(\ref{H eff(1)}), we can draw the conclusion that
\begin{eqnarray}
\label{H eff final}
H_{eff}&&=H_{eff}^{(0)}+\dfrac{1}{\hbar\omega}H_{eff}^{(1)} \nn\\
&&=\sum_k 
\begin{pmatrix}
a_k^\dagger & b_k^\dagger
\end{pmatrix}
\begin{pmatrix}
\mathcal{H}_{eff,11}(\vec{k}) & \mathcal{H}_{eff,12}(\vec{k}) 
\\ \mathcal{H}_{eff,21}(\vec{k}) & \mathcal{H}_{eff,22}(\vec{k})
\end{pmatrix}
\begin{pmatrix}
a_k \\ b_k
\end{pmatrix} \nn \\
\end{eqnarray}
where

\begin{eqnarray}
\label{ker11}
\mathcal{H}_{eff,11}&&=\left[-2t_{NNN}+\dfrac{8t_0t_1}{\hbar\omega}\dfrac{\mathcal{J}_1(\beta)^2}{\beta}\right]\sum_{j=1}^{3}\cos(\vec{k}\cdot\vec{v_j}) \nn \\
&&+\dfrac{12t_0t_1}{\hbar\omega}\dfrac{\mathcal{J}_{1}(\beta)^2}{\beta}
\end{eqnarray}
\begin{eqnarray}
\label{ker22}
\mathcal{H}_{eff,22}&&=\left[-2t_{NNN}-\dfrac{8t_0t_1}{\hbar\omega}\dfrac{\mathcal{J}_1(\beta)^2}{\beta}\right]\sum_{j=1}^{3}\cos(\vec{k}\cdot\vec{v_j}) \nn \\
&&-\dfrac{12t_0t_1}{\hbar\omega}\dfrac{\mathcal{J}_{1}(\beta)^2}{\beta}
\end{eqnarray}
and
\begin{eqnarray}
\label{ker12}
\mathcal{H}_{eff,12}&&=-t_0\mathcal{J}_0(\beta)\sum_{j=1}^{3}e^{-\mathbbm{i}\vec{k}\cdot\vec{u_j}}-\dfrac{d}{2}
\end{eqnarray}
\begin{eqnarray}
\label{ker21}
\mathcal{H}_{eff,21}&&=-t_0\mathcal{J}_0(\beta)\sum_{j=1}^{3}e^{\mathbbm{i}\vec{k}\cdot\vec{u_j}}+\dfrac{d}{2}
\end{eqnarray}

The $2\times2$ matrix in the above equation is the so-called kernel of Hamiltonian, denoted as $\mathcal{H}_{eff}$, which can be spaned by 2-ranked identity matrix and three 2D Pauli matrices according to linear algebraic theory. The final result is exactly Eq.(\ref{Heff}) in the article.

\section{The Berry phase and symmetry analysis of nodal-line phase}
\begin{figure}
	\includegraphics[width=0.5\textwidth]{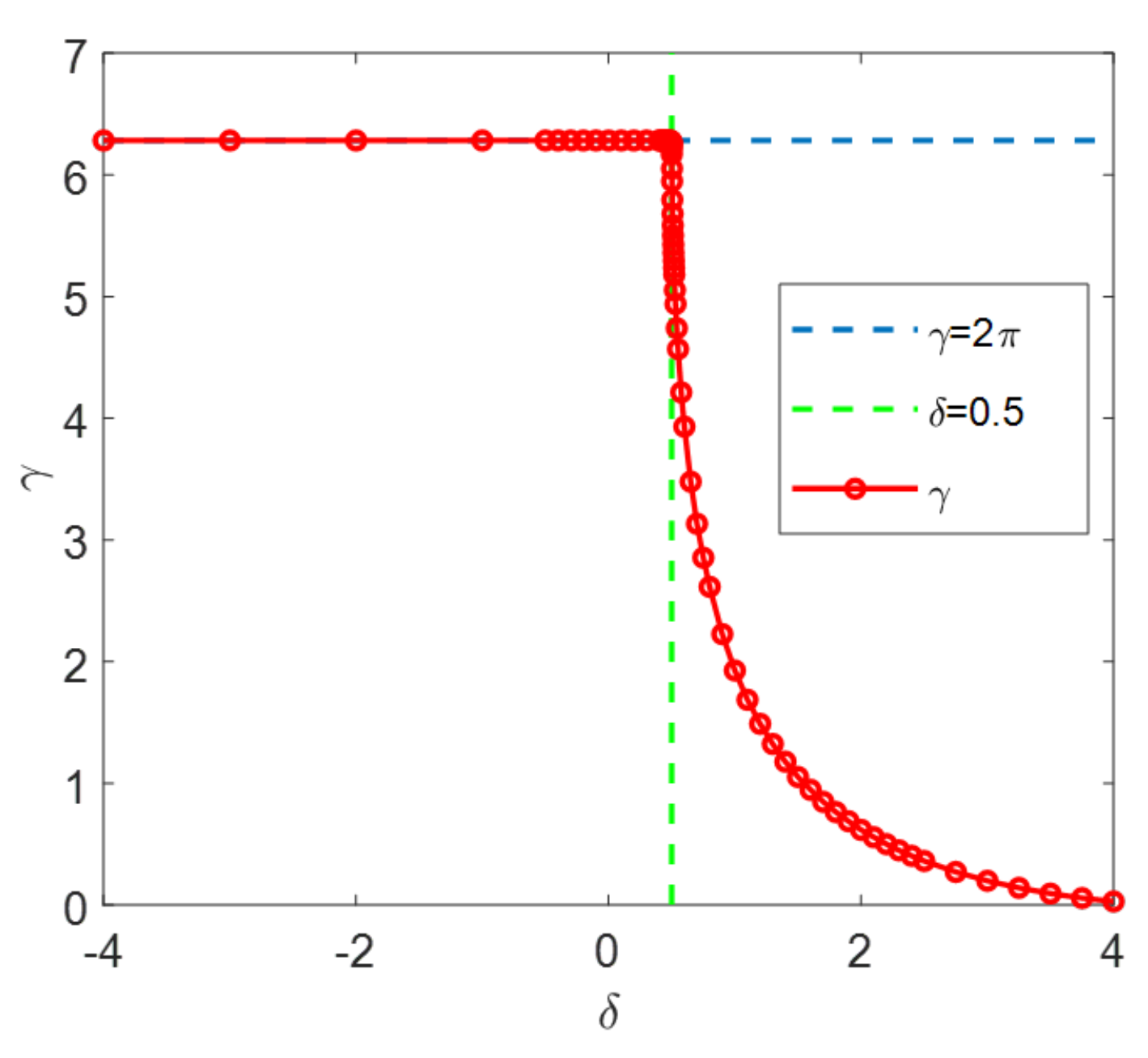}
	\caption{\textbf{The change of Berry phase with shaking factor $\delta$.} Here we set $\beta=2.4048$. When $\delta<0.5$, $\gamma$ is always equal to $2\pi$. When $\delta>0.5$, $\gamma$ continuously decreases from $2\pi$ to 0.
	}\label{fig:fig9}
\end{figure}

 The change of the Berry phase around the closed path in Fig.\ref{fig:fig5} with the change of $\delta$ is calculated and studied for fixed $\beta=2.4048$ in nodal-line phases. From Fig.\ref{fig:fig9}, we can observe that for all $\delta <0.5$, the Berry phase $\gamma$ is always equal to $2\pi$, and a mutation of the Berry phase occurs at $\delta=0.5$. The nodal-line in the first Brillouin zone at $\delta=0.5$ is shown in Fig.\ref{fig:fig8}(b3) by red dashed hexagon. For $\delta>0.5$, the area of nodal-line lies all inside the first Brillouin zone, so the nodal-line is a complete closed unrounded loop. However, when $\delta<0.5$, the solution $(k_x, k_y)$ satisfying $E_+-E_-=0$ will beyond the scope of the first Brillouin zone, so the nodal-line becomes incomplete. Hence $\delta=0.5$ point can be considered as a phase transition point between the complete nodal-line phase and incomplete nodal-line phase.
 
The well depth different $\delta$ is connected to the symmetry of the lattice. When $\delta =0$ point A is equivalent to point B, the lattice meets six-fold rotational symmetry. When lattice meets six-fold rotational symmetry, nodal lines degenerate into points at the six vertices of the Brillouin zone. When $\delta >0$, the six-fold rotational symmetry breaks and nodal lines appear. If we fixed $\beta$ at 2.4048, the nodal-line phase is present for a range of parameter $\delta$, which is $\delta\in(0,4.5)$. There is an interesting phenomenon that the situations for $\delta >0$ and $\delta <0$ are asymmetric. One possible reason is that in Eq.(\ref*{depth of A and B}) we artificially stipulate the initial state of point A and B, and it makes point A and B asymmetric.

\end{subappendices}

\bibliographystyle{apsrev}
\bibliography{my}

\end{document}